\documentclass[aps,prb,twocolumn,showpacs,superscriptaddress]{revtex4-1}
\pdfoutput=1
\usepackage{amsfonts} 
\usepackage{amsmath}
\usepackage{amssymb}
\usepackage{bm}
\usepackage[usenames, dvipsnames]{color} 
\usepackage{dcolumn}
\usepackage{epic} 
\usepackage{epsfig}
\usepackage{graphicx}
\usepackage[percent]{overpic}
\usepackage[lofdepth,lotdepth,caption=false]{subfig}

\usepackage[breaklinks,colorlinks = true,linkcolor = blue,urlcolor=blue,citecolor=blue]{hyperref}
\usepackage{soul}
\usepackage{times}
\usepackage{ulem}
\usepackage{wrapfig} 
\usepackage{xy} 

\newcommand{\beq}{\begin{equation}}
\newcommand{\eeq}{\end{equation}}
\newcommand{\beqa}{\begin{eqnarray}}
\newcommand{\eeqa}{\end{eqnarray}}

\newcommand{\Rmnum}[1]{\expandafter\@slowromancap\romannumeral #1@}

\newcommand{\mbz}{{\mathbb{Z}}}

\newcolumntype{M}[1]{>{\centering\arraybackslash}m{#1}}

\begin{document}

\title{Quantum Spin Liquid in a Breathing Kagome Lattice}


\author{Robert Schaffer}
\author{Yejin Huh}
\author{Kyusung Hwang}
\affiliation{Department of Physics and Center for Quantum Materials,
University of Toronto, Toronto, Ontario M5S 1A7, Canada.}
\author{Yong Baek Kim}
\affiliation{Department of Physics and Center for Quantum Materials,
University of Toronto, Toronto, Ontario M5S 1A7, Canada.}
\affiliation{Canadian Institute for Advanced Research, Quantum Materials Program,
Toronto, Ontario M5G 1Z8, Canada}
\affiliation{School of Physics, Korea Institute for Advanced Study, Seoul 130-722, Korea.}


\begin{abstract}
Motivated by recent experiments on the vanadium oxyfluoride material DQVOF, we examine possible spin liquid phases on a breathing kagome lattice of S=1/2 spins. By performing a projective symmetry group analysis, we determine the possible phases for both fermionic and bosonic $\mbz_2$ spin liquids on this lattice, and establish the correspondence between the two. The nature of the ground state of the Heisenberg model on the isotropic kagome lattice is a hotly debated topic, with both $\mbz_2$ and U(1) spin liquids argued to be plausible ground states. Using variational Monte Carlo techniques, we show that a gapped $\mbz_2$ spin liquid emerges as the clear ground state in the presence of this breathing anisotropy. Our results suggest that the breathing anisotropy helps to stabilize this spin liquid ground state, which may aid us in understanding the results of experiments and help to direct future numerical studies on these systems.
\end{abstract}
\date{\today}
\maketitle

\section{Introduction}

Quantum spin liquids, systems in which quantum fluctuations prevent the magnetic ordering of the spin degrees of freedom down to zero temperature, have been the source of much recent theoretical and experimental interest.\cite{Anderson1973,anderson1987resonating,Lee2008a,Balents2010} These systems are characterised by their long ranged entangled states, preserve symmetries down to zero Kelvin, and have been shown to exhibit fascinating properties such as a topological ground state degeneracy and fractionalised spin excitations.\cite{savary2016quantum} Experimentally, spin liquid physics has been studied in organic triangular lattice materials, and systems such as pyrochlore quantum spin ice and hyperkagome materials have been suggested as possible realisations of this physics.\cite{Shimizu2003,Yamashita2010,gingras2014quantum,okamoto2007spin,lawler2008topological,chen2008spin,lawler2008gapless,zhou20084,podolsky2009mott,podolsky2011a,chen2013b} Theoretically, exactly solvable models such as the Kitaev model offer us a route to explore this physics in a controlled fashion, and much recent effort has been made to realise this physics in strongly spin-orbit coupled systems.\cite{kitaev2006anyons,jackeli2009mott,chaloupka2010kitaev,plumb2014alpha,Takayama2014,Modic2014ch} The kagome antiferromagnet is of interest both experimentally and theoretically, with numerical studies\cite{jiang2008density,yan2011spin,depenbrock2012nature,jiang2012identifying,iqbal2011projected,iqbal2015spin,li2016z}
 suggesting a spin liquid\cite{sachdev1992kagome,wang2006spin,ran2007projected,hermele2008properties,lu2011z}
 as the ground state of the Heisenberg model, and the material Herbertsmithite believed to realise spin liquid physics at low temperatures.\cite{mendels2007quantum,Helton2007,imai2008cu,olariu200817,han2012fractionalized,pilon2013spin,fu2015evidence,potter2013mechanisms,huh2013optical,dodds2013quantum,punk2014topological}

Recent experiments on a vanadium based material [NH$_4$]$_2$[C$_7$H$_{14}$N][V$_7$O$_6$F$_{18}$] (diammonium quinuclidinium vanadium oxyfluoride; DQVOF) have suggested the possibility that this system exhibits spin liquid behaviour.\cite{aidoudi2011ionothermally,clark2013gapless} This material contains two-dimensional structures, consisting of three planes of vanadium ions (along with non-magnetic oxygen and fluorine ions) which are separated from other three-plane structures by non-magnetic atoms. These three planes consist of two kagome lattice planes of spin-1/2 V$^{4+}$ ions separated by one layer of spin-1 V$^{3+}$ ions arranged in a triangular lattice. The kagome lattices appear to have very little disorder and to be formed of equilateral triangles, which, along with the small spin-orbit coupling present for $3d$ atoms, suggests that the highly frustrated Heisenberg model may be a good approximation of the spin physics in these layers. However, the up and down triangles differ in size, requiring an anisotropic Heisenberg model to describe the physics of this breathing kagome lattice. These kagome layers appear to be well isolated from one another and to couple only weakly to the intermediate triangular lattice layers, offering a possible realisation of a nearly isolated kagome lattice antiferromagnet without disorder.

Motivated by this discovery, we study the anisotropic Heisenberg model on the kagome lattice. Classically, this model is equally frustrated to the fully isotropic Heisenberg model, with the ground state manifold being those states for which the sum of the classical spin vectors on each triangle is zero. However, the full quantum model may show substantial differences. The isotropic model has been studied using a number of methods, including variational Monte Carlo (VMC)\cite{iqbal2011projected,iqbal2015spin,li2016z} and density matrix renormalization group (DMRG)\cite{jiang2008density,yan2011spin,depenbrock2012nature,jiang2012identifying} numerical studies. The earlier VMC calculations indicated that a gapless U(1) spin liquid with a Dirac spinon excitation was the ground state.\cite{iqbal2011projected,iqbal2015spin} However, a more recent VMC calculation challenges this claim, finding instead that a gapped $\mbz_2$ spin liquid has the lower energy.\cite{li2016z} DMRG studies find a gapped spin liquid ground state for the nearest neighbour Heisenberg model.\cite{yan2011spin,depenbrock2012nature} When a next nearest neighbour Heisenberg interaction is added, the entanglement entropy can be calculated, which suggests a $\mbz_2$ spin liquid ground state.\cite{jiang2012identifying} However, the nearest neighbour limit is more subtle, with the numerical computations having difficulty unambiguously determining the nature of the ground state. This may be due to the presence of many energetically competing spin liquid states, or to the proximity to a quantum critical point. Hence, a perturbative deformation of the isotropic nearest neighbour model may serve to stabilize a unique spin liquid ground state. The breathing kagome lattice offers such an opportunity, which may lead to a better comparison between theoretical predictions and experimental results.

In this work, we investigate the quantum spin liquid ground state of the breathing kagome lattice. In particular, we show that the VMC in the fermion basis clearly favors a gapped $\mbz_2$ spin liquid. We narrow our search to consider only those phases which respect all of the symmetries of the Hamiltonian, as has been shown to be reliable for the isotropic kagome lattice Heisenberg model.\cite{li2016z} In order to systematically study these phases, we first classify the possible spin liquid phases, using the projective symmetry group (PSG) analysis of symmetric spin liquids.\cite{wen2002quantum} We make a clear connection between the results of our analysis and the isotropic kagome lattice results.\cite{wang2006spin,lu2011z} We also derive the connection between the bosonic and fermionic spin liquids, by considering the PSG of the vison excitations.\cite{Lu2014,Qi2015}

The remainder of the paper is organized as follows. In section \ref{sec:model}, we describe the details of the model, along with the symmetries present on this lattice. Next, we examine the possible symmetric spin liquid phases in this model using the PSG formalism. 
In section \ref{sec:fermion}, we consider the fermionic mean field states, and in section \ref{sec:boson} we consider the bosonic states. We also examine the relation between the two types of spin liquids in section \ref{sec:combined}. In addition to this, we comment on the relation between the isotropic kagome lattice and anisotropic kagome lattice solutions found, and show that each of the anisotropic kagome lattice solutions connects to certain isotropic lattice solutions.\cite{wang2006spin,lu2011z} Each isotropic lattice solution can also be considered as a special case of the anisotropic lattice solutions, as one would expect, fully describing the connection between the two PSGs. Following this, in section \ref{sec:VMC} we explore the energetics of the anisotropic kagome lattice model using the VMC technique. After describing the details of the calculation, we show our main result, that the gapped $\mbz_2$ spin liquid is the minimum energy variational solution. We conclude in section \ref{sec:disc} with a discussion of these results and of the possible relevance to the future numerical studies.


\section{\label{sec:model}Model and Symmetries}

We consider the spin-1/2 Heisenberg model on the anisotropic Kagome lattice. The Hamiltonian for this model takes the form
\begin{equation}
H = J_{\triangle}\sum_{\langle ij \rangle \in \triangle} \bm{S}_i \cdot \bm{S}_j + J_{\nabla}\sum_{\langle ij \rangle \in \nabla} \bm{S}_i \cdot \bm{S}_j
\end{equation}
where $J_\triangle$ and $J_\nabla$ are the strengths of the interactions on links in up and down triangles respectively, $\langle ij \rangle$ denotes sums over nearest neighbour sites and $\bm{S}_i$ is the spin-1/2 operator at site $i$. We restrict our study to the case in which $J_\triangle, J_\nabla >0$. Without loss of generality, we will take $J_{\triangle} > J_\nabla$ and set $J_{\triangle} = 1$.

This model respects a subset of the symmetries of the isotropic Kagome lattice. Both translational symmetries ($T_1, T_2$) and the time reversal symmetry ($T$) are present. The $C_6$ symmetry of the isotropic Kagome lattice is broken down to a $C_3$ subset. There are three inequivalent locations about which $C_3$ can rotate the system; we choose the center of rotation to be about the center of the hexagon, for consistency with the isotropic lattice. In addition, the reflection which ran parallel to the primitive lattice vector is not present in the anisotropic lattice, while the reflection ($\sigma$) which runs perpendicular to the lattice vectors remains. This lattice therefore has the space group symmetry p3m1.

\begin{figure}
\centering
\includegraphics[scale=1.2]{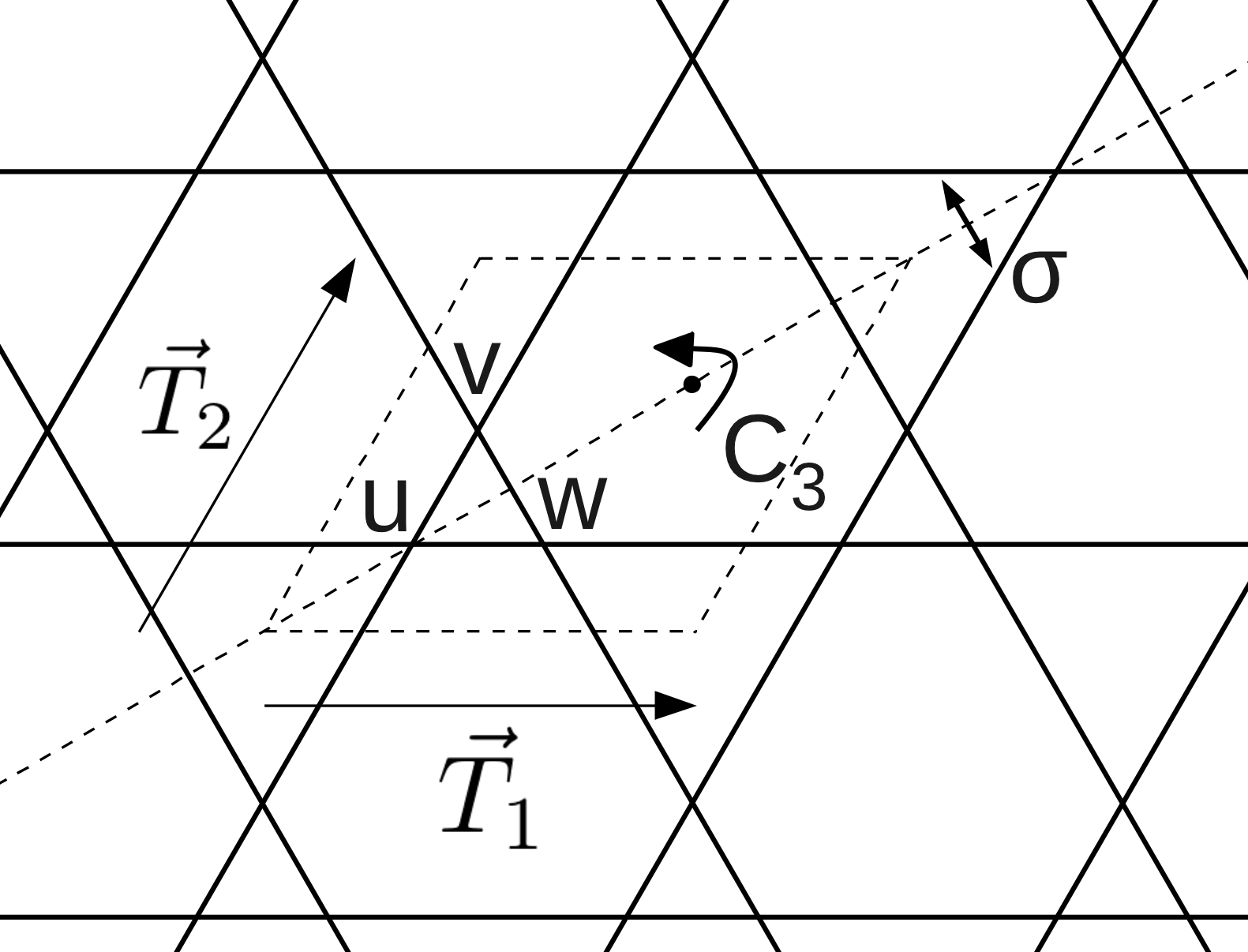}
\caption{The anisotropic (breathing) kagome lattice. Outlined is a single unit cell, consisting of three lattice sites. Also shown are the four symmetry operations required to generate the space group. The size difference between the inequivalent triangles has been exaggerated compared to experiment, for clarity of presentation.\cite{aidoudi2011ionothermally}}
\label{fig_lattice}
\end{figure}

\section{\label{sec:fermion}Fermionic Spin Liquid states}

Next, we will consider the fermionic slave-particle description of the spin degrees of freedom.\cite{PhysRevLett.58.2790,Wen2004B} In this theory, the spin operators are represented by a fermionic bilinear with the same commutation relations as the original spins. We choose
\begin{equation}
S_i^\mu = \frac{1}{2} f^\dag_{i\alpha}[\sigma^\mu]_{\alpha \beta}f_{i\beta},
\end{equation}
where $\sigma$ represents the Pauli matrices, $f_{i\alpha}(f_{i\alpha}^\dag)$ annihilates (creates) a fermion of type $\alpha$ on site $i$ and $\alpha,\beta$ $\in$ $\uparrow,\downarrow$. This representation of the spins in terms of fermions, known as spinons, along with the single fermion per site constraint
\begin{equation}
f_{i \uparrow}^\dag f_{i \uparrow} + f_{i \downarrow}^\dag f_{i \downarrow} = 1,
\end{equation}
gives a faithful representation of the Hilbert space. At this point, the analysis is exact (i.e. no approximations have been made) and the spin Hamiltonian appears as a quartic spinon Hamiltonian. Using a mean-field decoupling we can decompose this Hamiltonian in terms of auxiliary fields, leading to a quadratic Hamiltonian in terms of the fermionic spinons. 

To describe symmetric spin liquids which preserve the SU(2) spin symmetry of the underlying spin Hamiltonian, we choose a decomposition in which only the singlet spinon hopping and pairing channels appear in the mean field Hamiltonian. These take the form
\begin{equation}
\chi_{ij} = \langle f_{i\alpha}^\dag f_{j\alpha}\rangle, \quad \Delta_{ij} = \langle f_{i\alpha} [i\tau^2]_{\alpha \beta} f_{j\beta}\rangle,
\end{equation}
where $\tau$ represents Pauli matrices. This decomposition leads to a mean field Hamiltonian of the form 
\begin{align}
H_0 &= \sum_{ij}J_{ij}\vec{f}_i{}^\dag U_{ij} \vec{f}_j + \sum_i \sum_{l=1}^3 \Lambda^l \vec{f}_i{}^\dag \tau^l \vec{f}_i \nonumber \\
U_{ij} &= \begin{bmatrix}
\chi_{ij}^\dag & \Delta_{ij} \\
\Delta_{ij}^\dag & -\chi_{ij}
\end{bmatrix}, \qquad \vec{f}_i = \begin{bmatrix}
f_{i \uparrow} \\
f_{i \downarrow}^\dag
\end{bmatrix}.
\end{align}
The terms $\Lambda^l$ are Lagrange multipliers which enforce the single fermion per site constraint on average. In order to enforce this physical requirement on each site individually, a projection is required, which is done in section \ref{sec:VMC}.

This mean-field Hamiltonian has an SU(2) gauge freedom originating from the slave-particle representation of the spin operators. To be specific, the physical spin operators and physical Hilbert space are invariant under the local SU(2) gauge transformation
\begin{align}
\vec{f}_i &\rightarrow W_i^\dag \vec{f}_i
\end{align}
where the $W_i$ are SU(2) matrices. The corresponding ansatz transforms as
\begin{align}
U_{ij} &\rightarrow W_i U_{ij} W_j^\dag,
\end{align}
Although the mean-field ansatz changes its form under the transformation, the physical spin state corresponding to the ansatz remains unchanged.

We will restrict our focus to symmetric spin liquids, in which the symmetries of the Hamiltonian are realized by the physical spin wavefunction. The corresponding mean field wavefunction, however, may change under a symmetry transformation, as long as the transformed wavefunction is gauge equivalent to the original; such wavefunctions are invariant under the symmetry once projected. Mean field ansatz $U_{ij}$ which have such a wavefunction as the ground state are invariant under a combined symmetry and gauge transformation, such that the equation
\begin{align}
\label{eq_transform}
U_{ij} = G_{S}(i) U_{S^{-1}(i) S^{-1}(j)}G_S^\dag (j),
\end{align}
holds on all combinations of sites $i$ and $j$.\cite{wen2002quantum,lu2011z} Here $S$ is a symmetry group transformation and $G_S$ is the corresponding gauge transformation, which is site dependent in general.

These transformations $G_S$, along with the corresponding symmetry transformations, define the projective symmetry group (PSG) of an ansatz $U_{ij}$.\cite{wen2002quantum} The PSG describes the transformations under which an ansatz is left unchanged, offering a tool for distinguishing different states with the same physical symmetries. Under a gauge transformation $W_i$ the transformations $G_S$ transform according to $G_S(i) \rightarrow W_i G_S(i) W^\dag_{S^{-1}(i)}$. We can therefore determine groups of equivalent $G_S$'s which can be connected by gauge transformations, and use this to distinguish different quantum states.

For a given ansatz $U_{ij}$, we can determine the set of all gauge transformations which leave the ansatz invariant, i.e., those gauge transformations with the property that $U_{ij} = W_i U_{ij} W_j^\dag$. This set forms a group which takes a role of particular importance in our analysis, and is known as the invariant gauge group (IGG) of the ansatz.\cite{wen2002quantum} Any nontrivial set of symmetry operations whose product is the identity leads to a set of gauge transformations which must multiply to an element of the IGG. For an example of this, we consider the commutation relation between the translation operators, $T_2^{-1} T_1^{-1} T_2 T_1 = I$. Acting these operations on the ansatz in turn, combined with their gauge transformations, we can see that $T_2^{-1}G_{T_2}^\dag T_1^{-1}G_{T_1}^\dag G_{T_2} T_2 G_{T_1} T_1$ must act trivially on the ansatz (or leave the ansatz invariant). Hence, the gauge transformation portion of this must belong to the IGG, i.e.
\begin{equation}
G_{T_2}^\dag(T_1^{-1}(i))G_{T_1}^\dag(i)G_{T_2}(i)G_{T_1}(T_2^{-1}(i)) = W'_i
\end{equation}
where $W'_i$ is an IGG transformation. 

In this paper, we will consider possible ans\"atze whose IGG is $\mbz_2$, i.e. the product of gauge matrices described above must be $\pm 1$. The gauge transformations defined above must therefore satisfy $W'_i = \eta_{12} I$, where I is the identity matrix and $\eta_{12} = \pm 1$.  Ans\"atze which satisfy $\eta_{12} = +1$ and $\eta_{12} = -1$ cannot be continuously connected to one another. These $\mbz_2$ variables thus distinguish different quantum phases. By solving these equations for each relation between the symmetry operations, i.e. all of the relations in a presentation of the space group, we can categorize the possible quantum phases to which an ansatz can belong.

\begin{table*}

\begin{center}
{\renewcommand{\arraystretch}{1.5}
\renewcommand{\tabcolsep}{0.2cm}
	\begin{tabular}{|c|ccccc|cc|ccc|}
		\hline
		No. & $\eta_T$ & $\eta_{\sigma T}$ & $\eta_\sigma$ & $\eta_{C_3}$ & $\eta_{12}$ & $g_\sigma$ & $g_{C_3}$ & $\Lambda$ & n.n. & n.n.n. \\ \hline
		1,2 & -1 & 1 & 1 & -1 & $\pm$ 1 & $\tau_0$ & $\tau_0$ & $\tau_1$,$\tau_3$ & $\tau_1$,$\tau_3$ & $\tau_1$,$\tau_3$ \\ \hline
		3,4 & -1 & 1 & -1 & -1 & $\pm$1 & $i\tau_2$ & $\tau_0$ & 0 & 0 & $\tau_1$,$\tau_3$ \\ \hline
		5,6 & -1 & -1 & -1 & -1 & $\pm$1 & $i\tau_3$ & $\tau_0$ & $\tau_3$ & $\tau_3$ & $\tau_1$,$\tau_3$ \\ \hline
	\end{tabular}}
\end{center}
\caption{The possible fermionic $\mbz_2$ spin liquid states. Shown are the associated quantum numbers ($\eta_T - \eta_{12}$), sublattice portions of the gauge transformation matrices ($g_{\sigma}$ and $g_{C_3}$) and allowed chemical potential terms ($\Lambda$) and bond amplitudes on nearest neighbour and next nearest neighbour bonds (n.n. and n.n.n.). The odd (even) numbered PSGs have $\eta_{12} = +1$ ($-1$).}
\label{tab_psgsol}
\end{table*}

As outlined in appendix \ref{app:PSG}, we find a total of 6 solutions to the PSG equations for the anisotropic kagome lattice which allow non-zero amplitudes for the $U_{ij}$. The form of these solutions are detailed in table \ref{tab_psgsol}. The full gauge transformation matrices take the form
\begin{align}
G_{T_1}(x,y,s) &= \eta_{12}^yI, \\
G_{T_2}(x,y,s) &= I, \\
G_{\sigma}(x,y,s) &= \eta_{12}^{xy} g_{\sigma}, \\
G_{C_3}(x,y,s) &= -\eta_{12}^{y(y+1)/2+xy} I,
\end{align}
where the position $(x,y,s)$ is represented by using the coordinate system in Fig. \ref{fig_lattice} and the sublattice index $s$. In this expression, $g_\sigma \equiv G_\sigma(0,0,s)$ can be chosen to be identical on each sublattice, and $G_{C_3}$ is proportional to the identity matrix. 

We have found 6 possible PSGs for the anistotropic kagome lattice, compared to 20 for the isotropic version of this lattice (Ref. \onlinecite{lu2011z}). In order to reconcile this result, we note that, on the anisotropic kagome lattice, up and down triangles are no longer related by symmetry. As a result, we have additional mean field parameters corresponding to these inequivalent bonds. Although we have fewer PSGs, we can realize all 20 of the isotropic kagome lattice PSGs as special cases of the 6 anisotropic kagome lattice PSGs.

The simplest way to see this is to consider the product of the gauge transformations associated with the symmetry transformations. Following the work of Lu {\it et. al.} in Ref. \onlinecite{lu2011z}, and considering that $C_3$ is $C_6^2$, in terms of the gauge transformations the solution for the isotropic kagome lattice case can be seen as a special point in the solution set for the anisotropic kagome lattice if $G_{C_3}(i) = G_{C_6}(i)G_{C_6}(C_6^{-1}(i))$ for all $i$, up to gauge transformations. This is the case due to the fact that in the isotropic case we have 
\begin{align}
U_{ij} &= G_{C_6}(i) U_{C^{-1}_6(i), C^{-1}_6(j)}G_{C_6}^\dag(j) \nonumber \\
&= G_{C_6}(i)G_{C_6}(C_6^{-1}(i)) U_{C^{-1}_3(i), C^{-1}_3(j)}G_{C_6}^\dag(C_6^{-1}(j))G_{C_6}^\dag(j),
\end{align}
and thus for these to match, $G_{C_6}(i)G_{C_6}(C_6^{-1}(i))$ must be gauge equivalent to $G_{C_3}(i)$. When this occurs, the isotropic kagome lattice PSG can be continuously connected to the anisotropic kagome lattice PSG.

Applying this to the spin liquids on the isotropic kagome lattice, we determine the correspondence between the isotropic kagome lattice states and those on the anisotropic kagome lattice. It can be seen that all of the PSGs for the anisotropic kagome lattice have special points which correspond to isotropic kagome lattice PSGs. This must be determined explicitly; anisotropic lattice PSGs 1-4 each correspond to three isotropic lattice PSGs, while PSGs 5 and 6 each correspond to four isotropic lattice PSGs. The isotropic lattice PSGs which share the values of $\eta_{12}$ and $G_{\sigma}$ all belong to the same anisotropic lattice PSG, regardless of their values of $G_{C_6}$.

As is noted in table \ref{tab_psgsol}, only PSGs 3 and 4 do not allow any amplitude for mean field parameters on nearest neighbour bonds, nor do they support chemical potential terms. In this case, only two free parameters appear up to second neighbour terms. For PSGs 5 and 6, nearest neighbour pairing and pairing chemical potential terms are disallowed, and second neighbour hopping and pairing are required to stabilize a $\mbz_2$ spin liquid state. In PSGs 1 and 2, both nearest neighbour hopping and pairing are allowed, as well as two on site chemical potential terms and second neighbour hopping and pairing. 

Of particular note for our considerations is PSG 2, which contains the $\mbz_2[0,\pi]\beta$ phase of the isotropic kagome lattice\cite{lu2011z} in the limit that the nearest neighbour hopping and pairing amplitudes are equal on the up and down triangles. On the anisotropic kagome lattice, we can remove one of the pairing parameters using a gauge transformation; here we choose to remove the nearest neighbour pairing on one of the triangles, $\Delta_{\triangle}$. The corresponding mean field Hamiltonian takes the form
\begin{align}
H_{MF} &= \sum_i \left[ \mu \sum_\alpha f_{i,\alpha}^\dag f_{i,\alpha} + \eta(f_{i,\uparrow}f_{i,\downarrow} + H.c.) \right] \nonumber\\
 &+ \chi_\triangle \sum_{\langle i,j\rangle \in \triangle ,\alpha} f_{i,\alpha}^\dag f_{j,\alpha} \nonumber\\
 &+ \sum_{\langle i,j\rangle \in \nabla } s_{i,j} \left[ \chi_\nabla \sum_\alpha f_{i,\alpha}^\dag f_{j,\alpha} + \Delta_{\nabla}(f_{i,\uparrow}f_{j,\downarrow} + H.c.) \right] \nonumber\\
 &+ \sum_{\langle \langle i,j\rangle \rangle } \nu_{i,j} \left[ \chi_2 \sum_\alpha f_{i,\alpha}^\dag f_{j,\alpha} + \Delta_{2}(f_{i,\uparrow}f_{j,\downarrow} + H.c.) \right],
\end{align}
up to second neighbour terms, where $\mu$ and $\eta$ are the chemical potential terms, $\chi_{\triangle,\nabla,2}$ are hopping terms on up-triangle nearest neighbour bonds, down-triangle nearest neighbour bonds and second neighbour bonds, $\Delta_{\nabla,2}$ are pairing terms on down-triangle nearest neighbour bonds and second neighbour bonds, and $s_{ij}$ and $\nu_{ij}$ take values of $\pm 1$, fixed by a gauge choice.

\begin{figure}
\centering
\label{fig_psg2}
\includegraphics[scale=1.2]{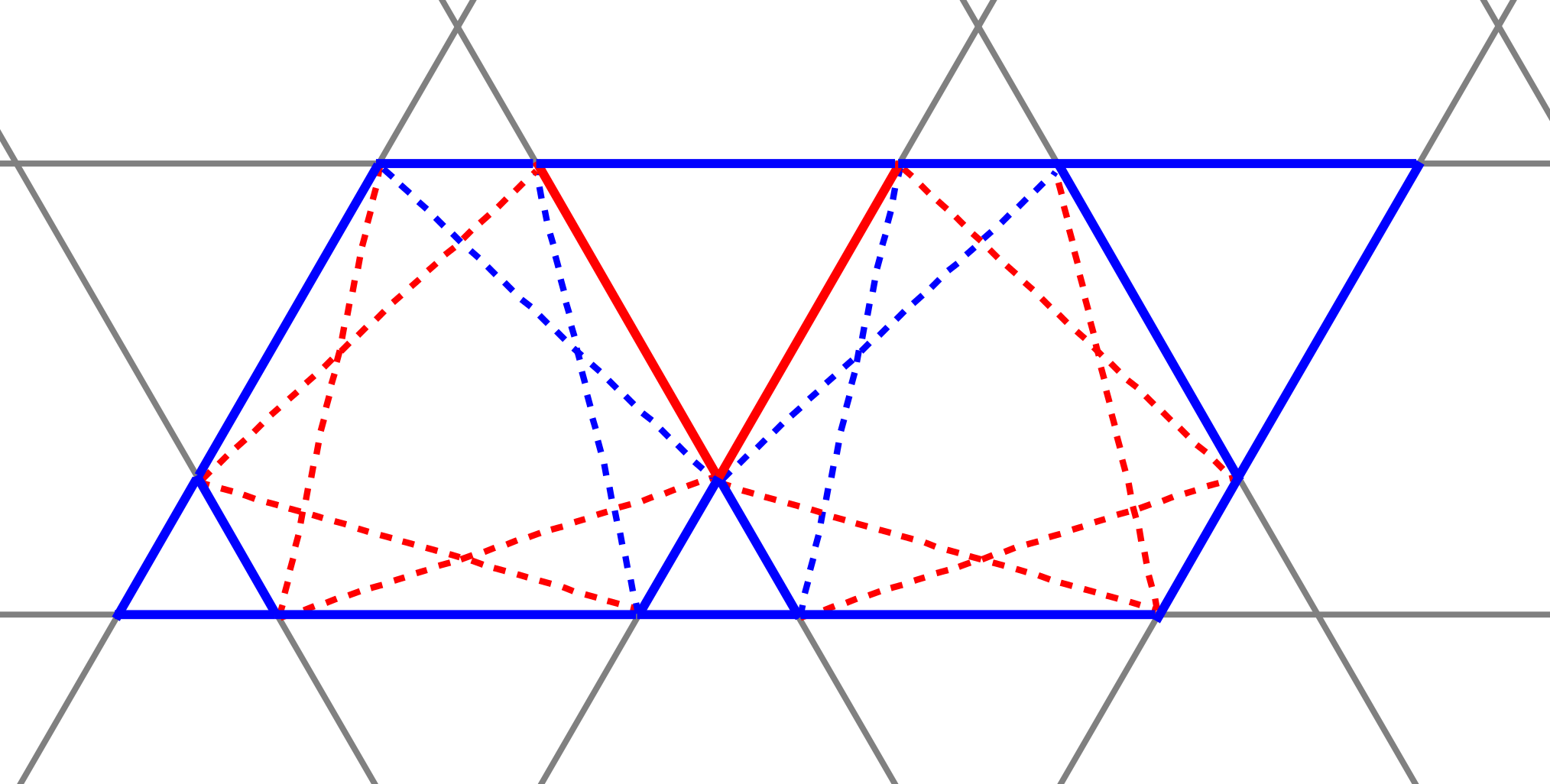}
\caption{The sign structure of the different bonds in PSG 2. Coloured is the extended unit cell which repeats over the lattice, with blue and red bonds representing $s_{ij} = +1$ and $s_{ij} = -1$ respectively on the down triangles. Also shown with dotted lines are the second neighbour bonds, with the same sign structure for $\nu_{ij}$ as above.}
\end{figure}

\section{\label{sec:boson}Bosonic Spin Liquid states}

We also consider the form of the bosonic spin liquid states which can appear on the anisotropic kagome lattice.\cite{sachdev1992kagome,wang2006spin} The analysis proceeds in a similar fashion to the fermionic case, where we now choose
\begin{equation}
S_i^\mu = \frac{1}{2} b^\dag_{i\alpha}[\sigma^\mu]_{\alpha \beta}b_{j\beta},
\end{equation}
where $\sigma$ again represents the Pauli matrices, $b_{i\alpha}(b_{i\alpha}^\dag)$ now annihilates (creates) a $boson$ of type $\alpha$ on site $i$ and $\alpha,\beta$ $\in$ $\uparrow,\downarrow$. The constraint on the number of spinons appearing on each site takes a different form between the two theories; in the Schwinger boson theory, we take that constraint to be
\begin{equation}
b_{i \uparrow}^\dag b_{i \uparrow} + b_{i \downarrow}^\dag b_{i \downarrow} = \kappa,
\end{equation}
where $\kappa=2S$ for the physical wavefunction of a spin system with spin $S$. In the mean-field theory, $\kappa$ is often taken to be a continuous positive real parameter.

Following Ref. \onlinecite{wang2006spin}, we note that the spin bilinears appearing in the Heisenberg Hamiltonian take the form
\begin{equation}
\vec{S}_i \cdot \vec{S}_j = :\hat{B}_{ij}^\dag \hat{B}_{ij} : - \hat{A}_{ij}^\dag \hat{A}_{ij},
\end{equation}
where the boson hopping and pairing operators $\hat{B}_{ij}$ and $\hat{A}_{ij}$ are of the form 
\begin{align}
\hat{B}_{ij} &= \frac{1}{2}(b_{i\uparrow}^\dag b_{j\uparrow} + b_{i\downarrow}^\dag b_{j\downarrow}), \\
\hat{A}_{ij} &= \frac{1}{2}(b_{i\uparrow} b_{j\downarrow} - b_{i\downarrow} b_{j\uparrow}).
\end{align}
This suggests a natural mean field Hamiltonian for the bosonic spinons, taking the form
\begin{align}
H_{0} &= \sum_{ij} J_{ij}(-A_{ij}^\ast \hat{A}_{ij} + B_{ij}^\ast \hat{B}_{ij} + h.c.) \nonumber\\
&- \mu \sum_i (\sum_\sigma b_{i\sigma}^\dag b_{i\sigma} - \kappa),
\end{align}
with $A_{ij}$, $B_{ij}$ and $\mu$ parameters which determine the form of the mean field wavefunction. Here, we have ignored constants which do not affect the form of the wavefunction for a given parameter set. Similar to the case of fermionic spinons, this Hamiltonian is invariant under a local gauge transformation which leaves the physical spin operators invariant; however, the gauge group of the transformations for bosons is U(1), rather than the SU(2) transformations for fermions. Under a gauge transformation, the bosons gain a phase factor
\begin{equation}
b_{i,\sigma} \rightarrow e^{i \phi_i}b_{i,\sigma}
\end{equation}
and the mean field parameters transform as 
\begin{equation}
A_{ij} \rightarrow e^{i(-\phi_i-\phi_j)} A_{ij}, ~~~B_{ij} \rightarrow e^{i(\phi_i-\phi_j)} B_{ij}.
\end{equation}

We can classify the possible bosonic spin liquid states which appear on this lattice in the same manner as was done for the fermionic spin liquids, keeping in mind the reduced space of possible gauge transformations. Because the gauge group is U(1), we can represent the gauge transformation matrices which define the transformation properties of the ansatz under the space group symmetries as $G_S = e^{i \phi_S}$, where $\phi_S \in [0,2\pi)$ are real numbers. Again analyzing spin liquids whose IGG is $\mbz_2$, a total of four bosonic spin liquids are found, which can be indexed by two $\mbz_2$ parameters $n_{12}$ and $n_{\sigma}$ $\in \{0,1\}$ as follows:
\begin{align}
\phi_{T_1}(x,y,s) &= \pi n_{12} y,\\
\phi_{T_2}(x,y,s) &= 0, \\
\phi_{\sigma}(x,y,s) &= \frac{\pi}{2} n_{\sigma} + \pi n_{12} xy, \\
\phi_{C_3}(x,y,s) &= \pi n_{12} xy + \frac{\pi}{2} n_{12} y(y+1).
\end{align}

\begin{figure}
\centering
(a)[0Hex,0Rhom]\\
\vspace{1mm}
\includegraphics[scale=1.0]{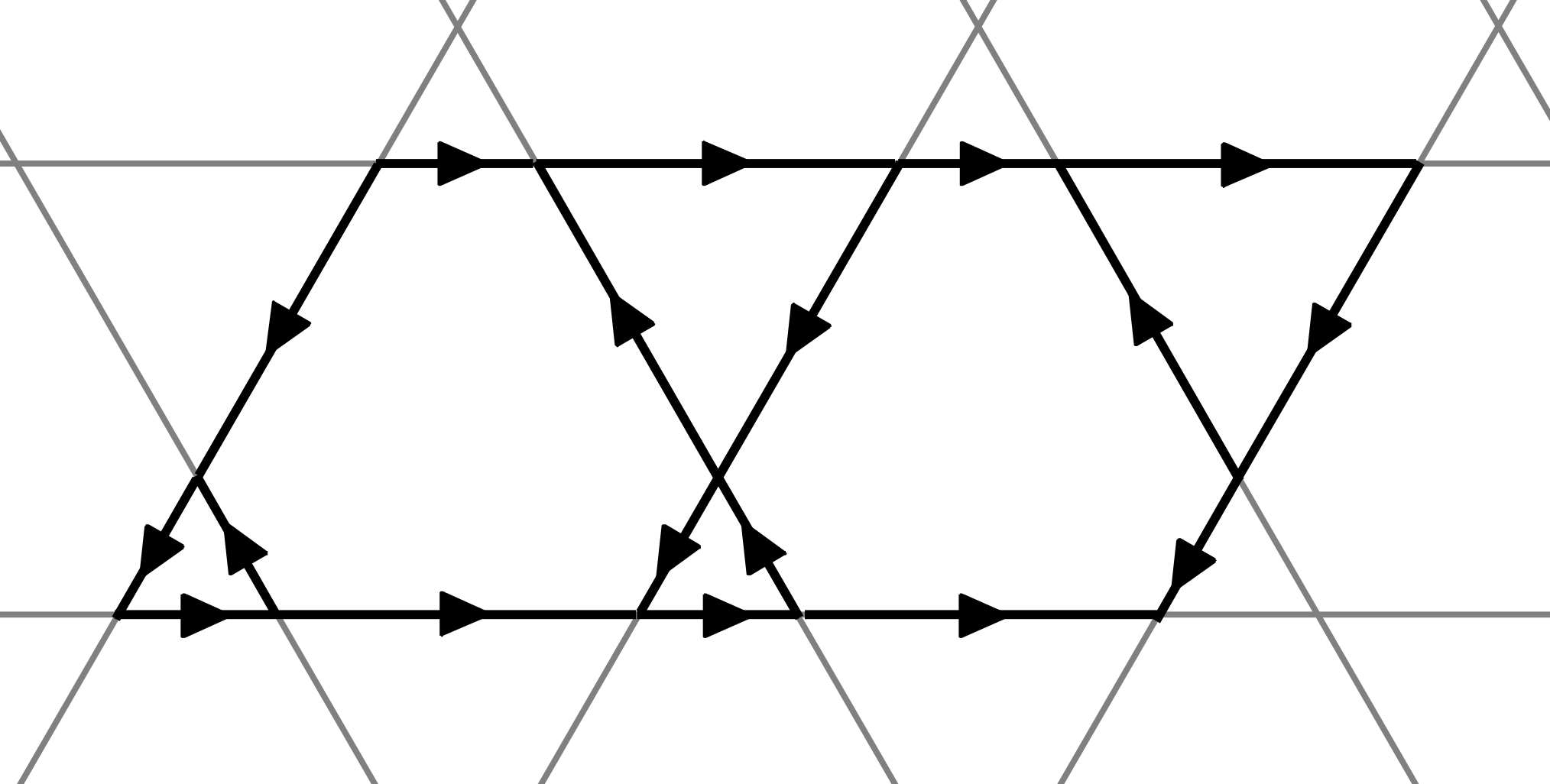}
\\ \vspace{2mm}
(b)[$\pi$ Hex,0Rhom]\\
\vspace{1mm}
\includegraphics[scale=1.0]{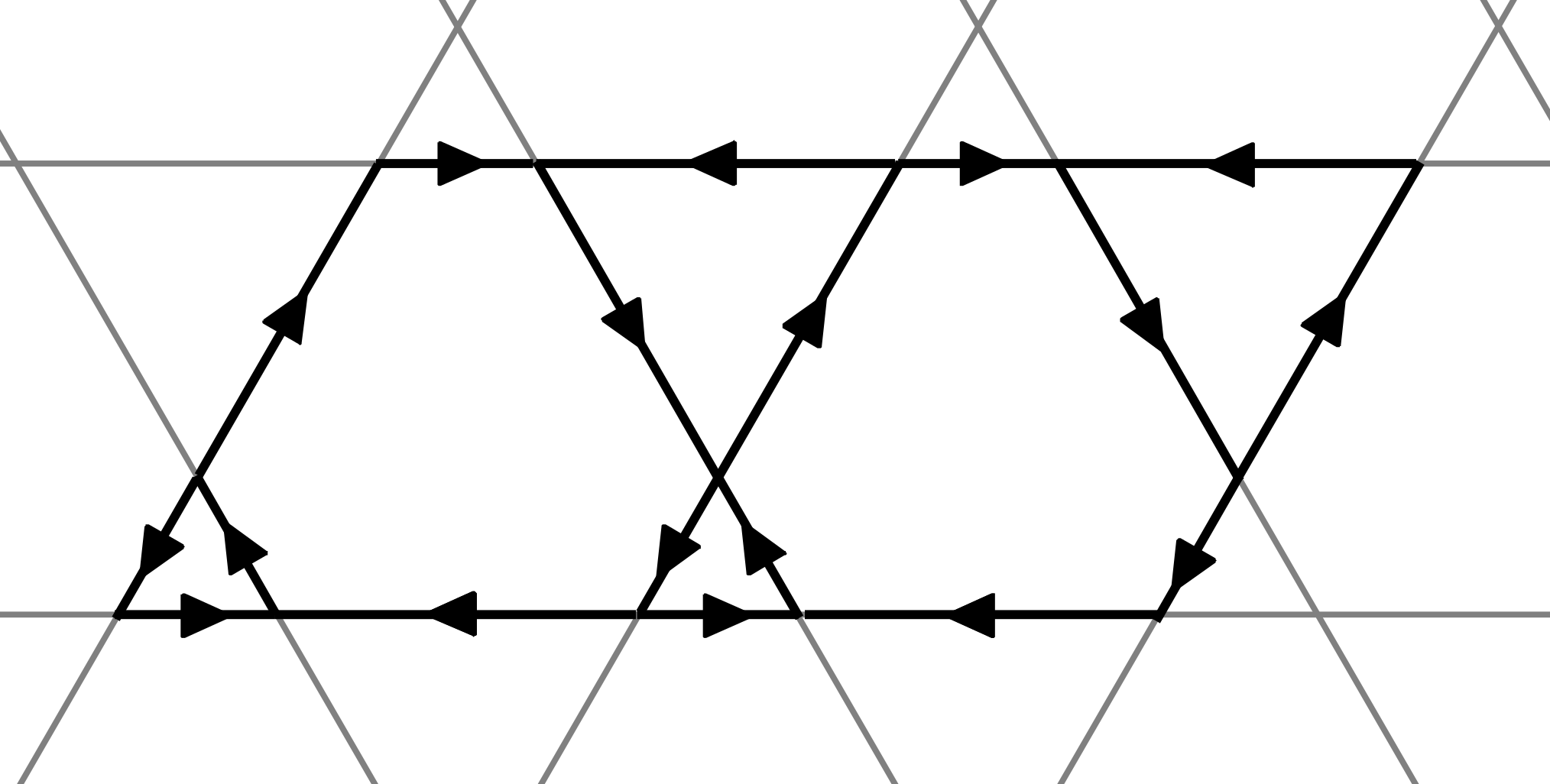}
\\ \vspace{2mm}
(c)[0Hex,$\pi$ Rhom]\\
\vspace{1mm}
\includegraphics[scale=1.0]{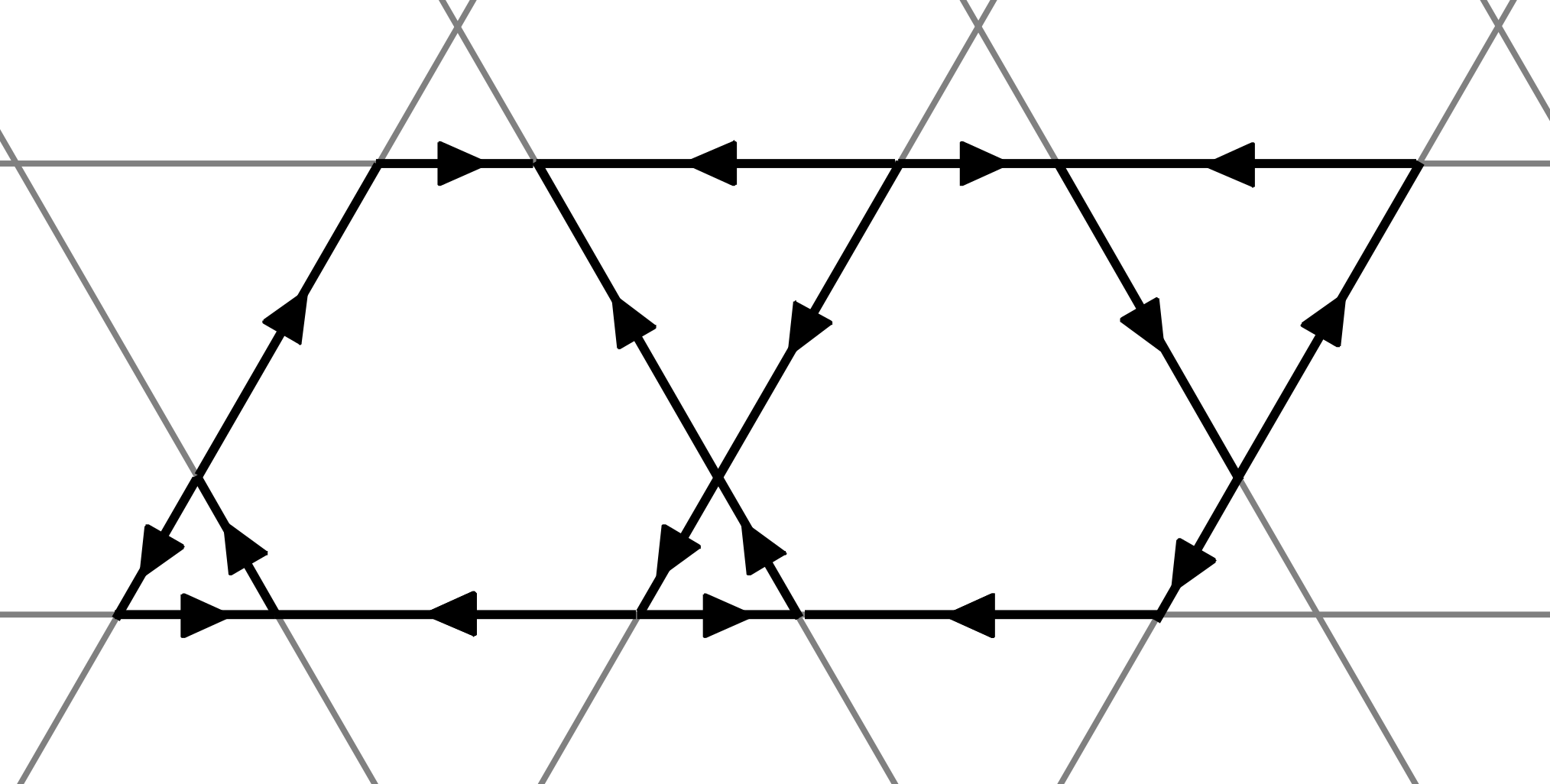}
\\ \vspace{2mm}
(d)[$\pi$ Hex,$\pi$ Rhom]\\
\vspace{1mm}
\includegraphics[scale=1.0]{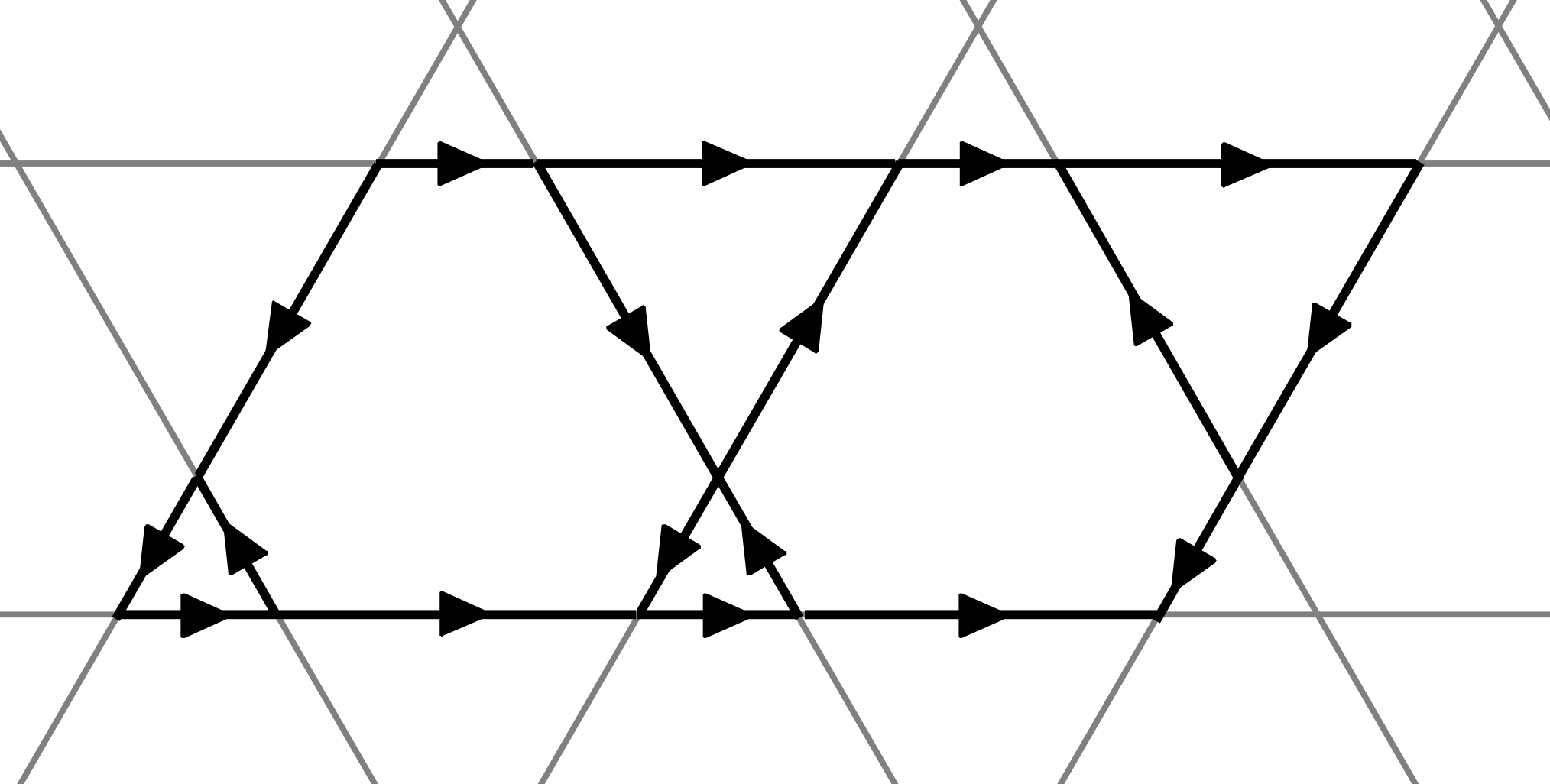}
\caption{The sign structure of the bosonic ansatz. States represented in (a) and (b) are in the same PSG, as are states represented in (c) and (d). These can be transformed into one another by changing the signs on all bonds on one triangle. This illustrates the reduction of the number of possible PSGs from the isotropic to anisotropic lattices.}
\label{fig_bos00}
\end{figure}

Notably, the number of PSGs for the bosonic states is again reduced from the number which appeared on the isotropic kagome lattice.\cite{wang2006spin} In this case, the correspondance between the two can be understood in a straightforward fashion, by comparing the fluxes which pass through the different loops on the lattice. When $n_{\sigma}$ is 0, the spin liquids have no nearest neighbour pairing, so here we will consider only $n_\sigma = 1$. Depending on whether $n_{12}$ is $0$ or $1$, we find that $0$ or $\pi$ flux passes through the length-8 rhombus. However, the flux passing through the length-6 hexagon is not fixed by the PSG, and will vary depending on the relative signs of the pairing parameters on the different triangles. Therefore, two phases which belonged to separate PSGs on the isotropic kagome lattice ([0Hex,0Rhom] and [$\pi$Hex,0Rhom], as well as [0Hex,$\pi$Rhom] and [$\pi$Hex,$\pi$Rhom]) now belong to the same PSG, as shown in Fig. \ref{fig_bos00}. As such, these phases will be referred to as [0Rhom] and [$\pi$Rhom] henceforth.

\section{\label{sec:combined}Mapping between fermionic and bosonic spin liquid states}

Correspondence between fermionic and bosonic $\mbz_2$ spin liquids can be found using symmetry fractionalization and fusion rules between fractionalized excitations.\cite{Lu2014,Qi2015} Since both spinons and visons are coupled to the emergent gauge field and transform projectively under different symmetries, the action of the symmetry operations on each fractionalized excitation is not gauge invariant. Hence it is useful to consider the gauge-invariant phase factor that the fractionalized excitations acquire after going through a series of transformations that, combined, are equivalent to the identity. Such transformations are listed in table~\ref{tab:identities}. From these analyses, one can find the symmetry fractionalization quantum numbers that characterize $\mbz_2$ spin liquids. Consider the relation 
\begin{equation}
f = b \times v \;,
\end{equation}
where $f$ is the fermionic spinon, $b$ the bosonic spinon, and $v$ the vison. Here
a fermionic spinon can be treated as a bound state of a bosonic spinon and a vison.\cite{Kitaev2003} This gives a relation between the phases accumulated among the three different particles, namely $\phi_f, \phi_b, \phi_v$ for the fermionic spinon, bosonic spinon, and vison, respectively.
As described in Ref.~\onlinecite{Lu2014}, two different relations are possible depending on the symmetry operation; 
$e^{i\phi_b} = e^{i\phi_f} e^{i\phi_v}$ is called the trivial fusion rule, 
where as $e^{i\phi_b} = -e^{i\phi_f} e^{i\phi_v}$ is called the non-trivial fusion rule. 
The extra minus sign in the non-trivial fusion rule comes from the mutual semionic statistics between spinons and visions.

Here we will go over the vison PSG first, discuss fusion rules, and then complete the mapping between the bosonic and fermionic spin liquid states. 

\subsection{Vison PSG}
The PSG for the visons can be found by describing them as pseudo-spins on a fully frustrated transverse field Ising model on the dual dice lattice. The transformations can be represented by the following matrices using the soft spin approach presented in Ref.~\onlinecite{Huh2011}, 
\begin{gather}
T_1 = \frac{1}{\sqrt{3}} \left( \begin{array}{cccc} 
	-e^{i 5\pi/6} 		& i \sqrt{2} 	& 0 	& 0 \\
	-\sqrt{2} e^{i \pi/6} 	& e^{i 5\pi/6} 				& 0 	& 0 \\
	0 				& 0 			& e^{i\pi/6} 		& -i \sqrt{2}  \\
	0 				& 0 			& \sqrt{2} e^{i 5 \pi/6}		& -e^{i\pi/6}
	\end{array}	\right),	\\
T_2 = \frac{1}{\sqrt{3}} \left( \begin{array}{cccc} 
	e^{i\pi/6} & \sqrt{2} e^{i 5 \pi/6} & 0 & 0 \\
	-i \sqrt{2}  & -e^{i\pi/6} & 0 & 0 \\
	0 & 0 & e^{-i\pi/6} & \sqrt{2} e^{-i 5 \pi/6} \\
	0 & 0 & i \sqrt{2} & -e^{-i\pi/6}
	\end{array}	\right),	\\
\sigma = \left( \begin{array}{cccc}
	0 & 0 & -1 & 0 \\
	0 & 0 & 0 & -1 \\
	-1 & 0 & 0 & 0 \\
	0 & -1 & 0 & 0 
	\end{array} \right), ~~~~
C_3 = \left( \begin{array}{cccc}
	-e^{i \pi/3} & 0 & 0 & 0 \\
	0  & 1 & 0 & 0 \\
	0 & 0 & e^{i 2 \pi/3} & 0 \\
	0 & 0 & 0 & 1
	\end{array} \right). 
\end{gather}
This is consistent with what was found in Ref.~\onlinecite{Hwang2015} with the addition of $T_1 = C_3 T_2^{-1} C_3^{-1}$. 

From these, the algebraic identites for the visons can be easily derived. The results are listed in table~\ref{tab:identities}. Note here that the sign of the $C_3^3$ operation has been chosen using the gauge freedom, {\it i.e.} $C_3 \rightarrow -C_3$, combined with flipping all the pseudo-spins of the dual dice lattice. (For example, if using the gauge choice made in Ref.~\onlinecite{Huh2011}, the lattice is explicitly $C_3$ symmetric. However, flipping all the spins under a $C_3$ operation is also valid, and will give $C_3 \rightarrow -C_3$.)

\begin{table}[h]
\centering
\begin{tabular}{c|c|c|c}
Algebraic Identities &  fermionic $f$ & bosonic $b$ & vison $v = b \times f$ \\
\hline \hline
$\sigma^2$ 					& $\eta_\sigma$ & $(-1)^{n_\sigma}$ & $1$  \\[1mm]
$T_2^{-1} T_1^{-1} T_2 T_1$	& $\eta_{12}$	& $(-1)^{n_{12}}$ 	& $-1$ \\[1mm]
$C_3^{3}$						& -1	& $1$	& 	$1$   \\[1mm]
$\sigma^{-1}T_2^{-1}\sigma T_1$	& 1			& 1					& 1  \\[1mm]	
$\sigma^{-1}T_1^{-1}\sigma T_2$	& 1			& 1					& 1  \\[1mm]	
$C_3 \sigma C_3 \sigma$		& $\eta_\sigma$	& $(-1)^{n_\sigma}$				& 1  \\[1mm]
$C_3^{-1} T_2^{-1} T_1 C_3 T_1$ & 1 		& 1					& 1  \\[1mm]
$C_3^{-1} T_1 C_3 T_2$		& 1				& 1					& 1  \\[1mm]
$T_1^{-1} T^{-1} T_1 T$		& 1				& 1					& 1  \\[1mm]
$T_2^{-1} T^{-1} T_2 T$		& 1				& 1					& 1  \\[1mm]
$\sigma^{-1}T^{-1}\sigma T$	& $\eta_{\sigma T}$	& $(-1)^{n_\sigma}$			& 1 \\[1mm]
$C_3^{-1} T^{-1} C_3 T$			& 1				& 1				& 1   \\[1mm]
$T^2$						& -1			& -1				& 1   \\[1mm]	
\hline
\end{tabular}
\caption{The algebraic identites of the bosonic spinon, fermionic spinon, and vison PSGs. Using the gauge freedom, the symmetry fractionalization quantum numbers of $C_3^{3}$ for the bosonic spinon and vison can be set to 1. While $\eta_{C_3}$ can be set to $\pm 1$ by the gauge-freedom, the non-trivial fusion rule fixes it to $-1$ for fermionic spinon in correspondence with the bosonic spinon. Similarly, some of the other terms have been chosen using the gauge freedom, while keeping with the correct fusion rule. }
\label{tab:identities}
\end{table}

\subsection{Fusion rule}
All unitary symmetry operations that can be written as $X^2 = e$ obey the non-trivial fusion rule. \cite{Qi2015} This can be shown by considering a state $|\Psi\rangle = f_r^\dagger f_{X(r)}^\dagger |G\rangle $ with two fermionic spinons (and equivalently as two vison - bosonic spinon bound states $f_r^\dagger = b_r^\dagger v_r^\dagger$) related by the symmetry $X$. Here $f^{\dagger}_{X(r)} = X f^{\dagger}_r X^{-1}$ and $|G \rangle$ is the ground state. Under the symmetry operation $X$, the fermionic spinons at $r$ and $X(r)$ are exchanged, leading to an extra minus sign compared to the operation on the equivalent bosons,
\begin{eqnarray}
X|\Psi\rangle 
&=& (X f_r^\dagger X^{-1})(X f_{X(r)}^\dagger X^{-1})|G\rangle
\nonumber\\
&=& f_{X(r)}^\dagger X^2 f_r^\dagger X^{-2} |G\rangle
\nonumber\\
&=& e^{i \phi_f} f_{X(r)}^\dagger f_r^\dagger |G\rangle
\nonumber\\
&=& -e^{i\phi_f} |\Psi\rangle 
\nonumber\\
&=& e^{i\phi_v} e^{i\phi_b}  |\Psi\rangle 
\end{eqnarray}
In our list, this applies to $C_3 \sigma C_3 \sigma$ and $\sigma^2$. 

Following the arguments of Ref.~\onlinecite{Lu2014}, $\sigma^{-1}T^{-1}\sigma T$ can be shown to obey the non-trivial fusion rule. 
Notice first that the anti-unitary squared operator $(T \sigma)^2$ obeys the trivial fusion rule because the time reversal symmetry 
provides the complex conjugation on the phase factor. Then, in the relation
\begin{equation}
(T \sigma)^2 = ( \sigma^{-1}T^{-1}\sigma T ) \cdot T^2 \cdot \sigma^2 \;,
\end{equation}
$(T \sigma)^2$ and $T^2$ obey the trivial fusion rule while $\sigma^2$ obeys the non-trivial fusion rule. 
Therefore $\sigma^{-1}T^{-1}\sigma T$ must obey the non-trival fusion rule. 

\begin{figure}
\includegraphics{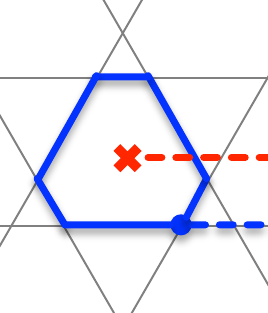} \quad\quad
\includegraphics{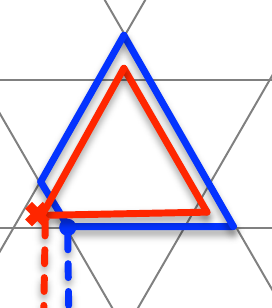}
\caption{Whether the series of operations that combine to identity loop a spinon (blue dot) around a vison (red x) can be observed by counting the number of crossings between the spinon and vison strings (blue and red dashed lines). The red and blue solid lines represent the symmetry operation that amounts to the identity.
The corresponding diagrams for $C_3^3$ (left) and $\sigma^{-1}T_2^{-1}\sigma T_1$ (right) are shown.}
\label{fig:fusion}
\end{figure}

The fusion rule for other operations can be seen by observing whether the operation effectively loops a spinon around a vison or not. 
In $Z_2$ spin liquids, the fractionalized excitations can be regarded as the end points of strings. As illustrated in Fig.\ref{fig:fusion},
we consider the fermionic string (blue dashed line) and vison string (red dashed line) \cite{Lu2014,chatterjee2016superconductivity}
associated with the fermionic spinon and vison. In the symmetry operation, when the fermionic string crosses the vision string, there
will be an extra minus sign. For $C_3^3$, this procedure is illustrated in Fig.~\ref{fig:fusion}, which shows that it obeys the non-trivial fusion rule. 
All the rest of the identity operations obey the trivial fusion rule. Fig.~\ref{fig:fusion} shows an example for $\sigma^{-1}T_2^{-1}\sigma T_1$. 

\subsection{Correspondence}

The analyses described in previous sections show that four out of the six fermionic spin liquid states have corresponding bosonic spin liquid states at the PSG level. However as mentioned in section \ref{sec:boson}, only two out of the four bosonic ans\"atze have weights on nearest neighbour bonds. Hence we conclude that only four out of the six fermionic spin liquid states are realized as the mean-field states (as shown earlier) and two of these four have corresponding bosonic spin liquid states. 
This correspondence has been summarized in table~\ref{tab:mapping}. 

\begin{table}
\centering
\begin{tabular}{M{0.7cm}|M{0.6cm}|M{0.6cm}|M{0.6cm}|M{0.6cm}|M{0.6cm} || M{0.6cm}|M{0.6cm}|c}
\hline
\multicolumn{6}{c||}{Abrikosov-fermion representation} & \multicolumn{3}{c}{Schwinger-boson representation} \\
\hline
No. & $\eta_T$ & $\eta_{\sigma T}$ & $\eta_{\sigma}$ & $\eta_{C_{3}}$	& $\eta_{12}$ & $n_{12}$ & $n_\sigma$ & label \\[1mm]
\hline
1	& -1 	& 1 	& 	1		& -1	& 	1	& 	1	& 	1	& 	$[\pi Rhom]$ 	\\[1mm]
2 	& -1 	& 1 	& 	1		& -1	& 	-1	& 	0 	& 	1	& 	$[0 Rhom]$ 	\\[1mm]
5	& -1 	& -1 	& 	-1		& -1	& 	1	& 	1 	& 	0	& 		\\[1mm]
6	& -1 	& -1 	& 	-1		& -1	& 	-1	& 	0 	& 	0	& 			\\[1mm]
\hline
\end{tabular}
\caption{Corresponding fermionic and bosonic spin liquid states. Schwinger-boson phases with $n_\sigma = 0$ have pairing amplitudes equal to zero on nearest neighbour bonds, and therefore do not have a well defined flux, as noted in section \ref{sec:boson}.}
\label{tab:mapping}
\end{table}


\section{\label{sec:VMC} Variational Monte Carlo Calculation for fermions}

Having examined the possible symmetric mean field states for this model, we would like to determine the lowest energy ansatz for the original spin model. In order to do so, we must first project the mean field wavefunction onto the space of physical spin wavefunctions, using the well-known Gutzwiller projection method.\cite{gros1989physics} By optimizing the energy with respect to our free mean field parameters, we find the best mean-field wavefunction for the spin state, and examine the properties of this ansatz.

\subsection{Details of the calculation}

We examined the energy of the projected wavefunctions
\begin{align}
|\Psi_{proj}\rangle &= P|\Psi_{\alpha_k}\rangle \\
P &= \prod_i (n_{i\uparrow} - n_{i\downarrow})^2
\end{align}
where $|\Psi_{\alpha_k}\rangle$ is the mean-field wavefunction for a given set of variational parameters $\alpha_k$, $P$ is the projector onto the physical spin Hilbert space and $n_{i\beta}$ is the number operator for $\beta$ spinons. The energy for a given state is computed through a Monte Carlo sampling over physical spin states, and the minimum energy state is computed using the stochastic reconfiguration (SR) method, which minimizes the energy with respect to the variational parameters.\cite{yunoki2006two,sorella2005wave} Calculations were performed for the fermionic spin liquid states, where methods involving determinants allow for an efficient calculation of expectation values of operators. We primarily use mixed periodic-antiperiodic boundary conditions, to avoid ambiguity for the calculation arising from sampling of the wavefunction at gapless points. The details of the numerical method are explained in appendix \ref{app:vmc}.


We focus primarily on the ansatz in PSG 2, which is continuously connected to the $\mbz_2[0,\pi]\beta$ phase. Examination of the energies of wavefunctions in other PSGs show consistently higher energies. We examine the mean field wavefunction with up to second neighbour non-zero amplitudes. Nearest neighbour hopping $\chi_{\triangle}$ is fixed to 1 ($\triangle$ denotes the up triangles), and we use the gauge freedom to set $\Delta_\triangle$ to 0. We thus have six free parameters in our minimization: on-site chemical potentials $\mu$ and $\eta$, nearest neighbour hopping and pairing $\chi_{\nabla}$ and $\Delta_\nabla$ and second nearest neighbour hopping and pairing $\chi_2$ and $\Delta_2$.

Calculations were performed on a 768 site lattice (16x16x3). The wavefunction was considered minimized when the effective forces (i.e. the change in energy with respect to the variational parameters, calculated within the SR scheme) averaged out to zero over a large number of runs.

\subsection{Results}

A number of important differences appear between the ans\"atze for the isotropic and anisotropic kagome lattices. In particular, comparing the most general ansatz for the $\mbz_2[0,\pi]\beta$ phase and the ansatz for spin liquids in PSG 2, two additional parameters appear in PSG 2; namely, $\chi_\nabla$ and $\Delta_{\nabla}$, which encode the breaking of the rotational symmetry from $C_6$ down to $C_3$. By setting $\chi_\nabla = 1$ and $\Delta_\nabla = 0$, we recover the isotropic ansatz; this provides a useful upper limit to our energy per site of $E = -0.42872(J_\triangle + J_\nabla)/2$ found in previous studies of the isotropic kagome lattice, which found a U(1) ground state.\cite{iqbal2011projected} In addition, this provides an additional avenue for breaking of the U(1) symmetry down to $\mbz_2$; if any of the parameters $\Delta_\nabla$, $\Delta_2$ or $\eta$ are non-zero in the minimized ansatz, the ansatz represents a $\mbz_2$ state. More importantly, the anisotropy in the variational parameters and the presence of additional variational parameters is expected to change the ground state configuration of all parameters.

Our main result is that with the increase of anisotropy between the inequivalent triangles, the $\mbz_2$ phase becomes the clear ground state within the VMC calculation. We determine this using two complementary approaches; first, we determine directly the energetic minimum from the SR procedure on the full parameter space, and, secondly, we minimize the energies at various fixed values of $\Delta_2$, showing directly that the energetic minimum occurs for a non-zero value of $\Delta_2$.

\begin{figure}
\centering
\begin{overpic}[scale=.38]
{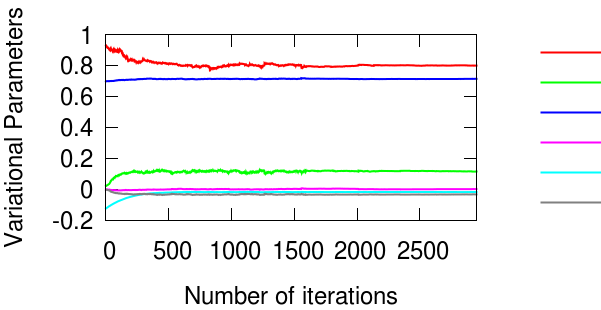}
\put(86.2,41.7){$\mu$}
\put(86.2,36.7){$\eta$}
\put(83.4,31.8){$\chi_\nabla$}
\put(82.6,26.5){$\Delta_\nabla$}
\put(84.1,21.8){$\chi_2$}
\put(83.35,16.5){$\Delta_2$}
\end{overpic}
\caption{A typical optimization run for the parameters, at anisotropy $J_{\nabla}/J_{\triangle}=0.7$. The fact that $\eta$ and $\Delta_2$ are non-zero implies that this is a $\mbz_2$ spin liquid.}
\label{fig_optimization}
\end{figure}
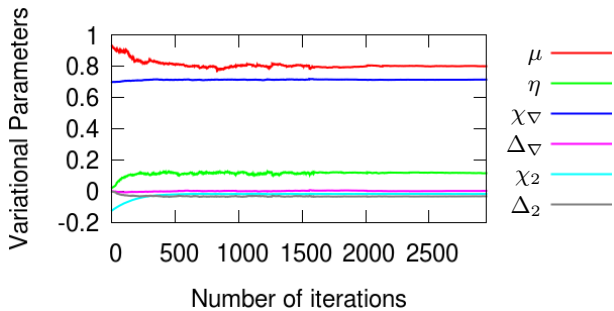

By considering the converged result of a large number of SR minimizations we can determine the values of the mean field parameters which we are varying. In the regime $0.5\leq J_{\nabla}/J_{\triangle}\leq0.8$, the variational parameters clearly converge to a single result, which yields a gapped wavefunction with a $\mbz_2$ gauge structure. Shown in Fig. \ref{fig_optimization} is a typical optimization run in the case $J_{\nabla}/J_{\triangle}=0.7$; here we see a clear convergence of the U(1) symmetry breaking parameters $\Delta_2$ and $\eta$ to non-zero values. For the converged order parameters, we find values of $\mu = 0.802(1)$, $\eta =0.116(5)$, $\chi_{\nabla} = 0.715(2)$, $\Delta_{\nabla} = 0.000(2)$, $\chi_2 = -0.0174(1)$ and $\Delta_2 = -0.0331(2)$.

The fact that $\Delta_{\nabla}$ appears to be zero (within error bars) is not determined by the projective symmetry group of the ansatz, but rather is a property of the solution. Interestingly, this is required in the $\mbz_2[0,\pi]\beta$ phase explored for the isotropic kagome lattice, suggesting that our minimum energy state is highly similar to the state found in the isotropic lattice. The ground state ansatz does not, however, directly correspond to this state, as the symmetry breaking between the hopping parameters on inequivalent triangles ($\chi_{\nabla}/\chi_{\triangle} = .715$) removes this possibility.

\begin{figure}
\centering
\includegraphics[scale=.35]{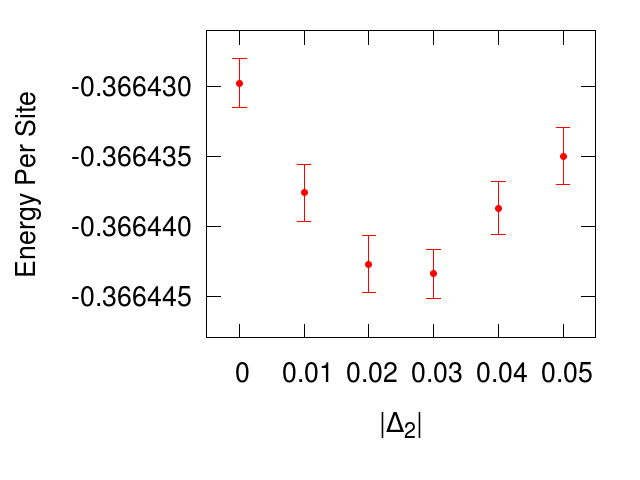}
\caption{The energy of the optimized wavefunction at a fixed value of $|\Delta_2|$, the pairing parameter on second neighbour bonds. Although the exact location of the minimum is difficult to obtain through this method, the minimum clearly occurs for a non-zero value of the pairing.}
\label{fig_engs}
\end{figure}

We next explore the energies of the ansatz with $\Delta_2$ fixed to a number of non-zero values, in order to directly confirm the breaking of the U(1) symmetry. Although the energy differences are small, they can be detected by performing a sufficiently large number of uncorrelated simulations. In Fig. \ref{fig_engs}, we show the minimum energy of the variational state as a function of $\Delta_2$, which confirms that the U(1) state is higher in energy than the minimum energy $\mbz_2$ state. A gauge symmetry relates positive and negative $\Delta_2$; as such, we plot this as a function of the absolute value of this parameter.

This result, that the ground state of the anisotropic kagome lattice is a $\mbz_2$ spin liquid, appears to be robust against the varying of the specific details of the calculation. The results shown are for a 768 site lattice with mixed boundary conditions at an anisotropy of $J_{\nabla}/J_{\triangle}=0.7$. Smaller lattice sizes were also explored, with 12x12x3, 8x8x3 and 4x4x3 lattices all showing results which are consistent with those found on the larger system size. For systems with anisotropy approaching the isotropic limit, numerical issues with multiple possible minima prevented convergence of the calculation. However, with anisotropy in the range listed ($0.5\leq J_{\nabla}/J_{\triangle}\leq0.8$), the convergence was clear and unambiguous. Finally, while periodic/periodic boundary conditions showed difficulty near the U(1) point due to degeneracy of the wavefunction, antiperiodic/antiperiodic boundary conditions give results which are fully consistent with mixed boundary conditions.


Our minimum energy state found can be compared directly to the energy of the best isotropic ansatz, as mentioned above. For the values $J_{\nabla}/J_{\triangle} = 0.7$, we find a minimum energy solution to have an energy of -0.366443(2)$J_\triangle$ per site. This has a lower energy than the isotropic ansatz evaluated at this value of the anisotropy, which gives an energy of -0.36441$J_\triangle$ per site.\cite{iqbal2011projected} This difference can be attributed to the anisotropy in the value of the nearest neighbour hopping parameters in the anisotropic ansatz, which favours states in which the nearest neighbour spin correlations are stronger on the stronger bonds.

\section{\label{sec:disc} Discussion}

The main effect that the addition of a breathing anisotropy appears to have on the nearest neighbour kagome lattice Heisenberg model is a stabilization of the $\mbz_2$ spin liquid state. On the isotropic kagome lattice, the nature of the ground state is controversial; different studies have found the presence of a U(1) and a $\mbz_2$ state, which appear to be extremely difficult to distinguish within the VMC technique.\cite{iqbal2011projected,li2016z} In our study, we also find a gap in the quasiparticle spectrum, which implies a gap for the spin excitations in the model.

While the anisotropy leads to the reduction of the number of spin liquids present in the PSG, more freedom is allowed in the choice of parameters present, leading to an overall increase in the set of possible wavefunctions which satisfy the symmetries of the lattice. Within these, we find a state which continuously connects to the $\mbz_2[0,\pi]\beta$ phase to be the energetically favoured ground state. 

The isotropic kagome lattice has been studied using other numerical methods, most notably the density matrix renormalization group (DMRG). Strong evidence of a $\mbz_2$ spin liquid ground state has been found in entanglement entropy measurements using this method; however, an additional second neighbour Heisenberg spin interaction for the unambiguous identification of this state, as the purely nearest neighbour model appears to be near a critical regime.\cite{jiang2012identifying} Our study suggests that adding a breathing anisotropy to the Hamiltonian may offer another path to stabilizing the $\mbz_2$ spin liquid phase, and therefore offers a worthwhile direction for further numerical studies. In particular, while our study indicates that a gapped spin liquid is the likely ground state for this model, estimating the magnitude of the singlet gap which may appear in experiments is a challenge which would require a DMRG study. Such a spin gap estimate could be compared to experimental results, offering evidence for this state being realized in materials such as DQVOF.
\medskip
\begin{acknowledgments}

We would like to thank Lucy Clark for bringing this material to our attention, as well as Yasir Iqbal, Hongchen Jiang and Lukasz Fidkowski for stimulating discussions. This research was supported by the NSERC, CIFAR and Centre for Quantum Materials at the University of Toronto.

\end{acknowledgments}

\bibliography{draft}

\appendix
\setcounter{equation}{0}
\setcounter{figure}{0}
\setcounter{table}{0}
\makeatletter
\renewcommand{\theequation}{A\arabic{equation}}
\renewcommand{\thefigure}{A\arabic{figure}}

\section{\label{app:PSG}PSG}

The anisotropic kagome lattice lies in the space group p3m1, which can be generated by four symmetries, \{$T_1 , T_2 , \sigma , C_3 $\}. A convenient presentation for this space group has eight relations between these generators,
\begin{gather}
\sigma^2 = e \\
T_2^{-1}T_1^{-1}T_2 T_1 = e \\
C_3^3 = e \\
\sigma^{-1}T_2^{-1}\sigma T_1 = e \\
\sigma^{-1}T_1^{-1}\sigma T_2 = e \\
C_3 \sigma C_3 \sigma = e \\
C_3^{-1}T_2^{-1}T_1 C_3 T_1 = e \\
C_3^{-1} T_1 C_3 T_2 = e.
\end{gather}

As described in the main text, the mean field Hamiltonian must be invariant under a combination of symmetry operations and gauge transformations, in order for the physical wavefunction to respect the symmetries. Considering sets of symmetry operations whose product is the identity leads to relations between the gauge transformations, which take the form
\begin{gather}
G_{\sigma}(\sigma(i))G_{\sigma}(i) = \eta_{\sigma} I, \label{eq_sig}\\
G_{T_2}^\dag(T_1^{-1}(i))G_{T_1}^\dag(i)G_{T_2}(i)G_{T_1}(T_2^{-1}(i))=\eta_{12} I, \label{eq_12}\\
G_{C_3}(C_3^{2}(i))G_{C_3}(C_3(i))G_{C_3}(i) = \eta_{C_3} I,\label{eq_S}\\
G_{\sigma}^\dag(T_2^{-1}(i))G_{T_2}^\dag(i)G_{\sigma}(i)G_{T_1}(\sigma(i)) = \eta_{\sigma T_1} I,\label{eq_sig1}\\
G_{\sigma}^\dag(T_1^{-1}(i))G_{T_1}^\dag(i)G_{\sigma}(i)G_{T_2}(\sigma(i)) = \eta_{\sigma T_2} I,\label{eq_sig2}\\
G_{C_3}(C_3 \sigma(i))G_{\sigma}(\sigma(i))G_{C_3}(i)G_{\sigma}(C_3^{-1}(i)) = \eta_{\sigma C_3} I,\label{eq_sigS}\\
G_{C_3}^\dag(T_2^{-1}T_1(i))G_{T_2}^\dag(T_1(i))G_{T_1}(T_1(i))\nonumber \\ 
\times G_{C_3}(i)G_{T_1}(C_3^{-1}(i))= \eta_{C_3 T_1}I,\label{eq_S1}\\
G_{C_3}^\dag(T_1(i))G_{T_1}(T_1(i))G_{C_3}(i)G_{T_2}(C_3^{-1}(i)) = \eta_{C_3 T_2}I\label{eq_S2}.
\end{gather}
Each of the $\eta$'s above take values $\pm 1$, as the IGG which we consider is $\mbz_2$.

The above relations which do not involve the rotation $C_3$ are equivalent to those found on the isotropic kagome lattice. The solution for these is therefore the same as on that lattice, which we will briefly recount here.

As with all two-dimensional lattices, we can perform a site dependent gauge transformation $W(i)$ which restricts the gauge transformations associated with the translational symmetries to
\begin{equation}
G_{T_1}(x,y,s) = \eta_{12}^y I, ~~~ G_{T_2}(x,y,s) = I.
\end{equation}
Here (and further on) $I$ denotes the 2x2 identity matrix for the fermionic transformations, and 1 for the bosonic transformations. Any future gauge transformations must preserve these choices, up to an overall factor of $\pm 1$ which has no effect on the ansatz $U_{ij}$.

The relations \ref{eq_sig1} and \ref{eq_sig2} restrict the gauge transformation $G_{\sigma}$ to take the form
\begin{align}
G_{\sigma}(0,y,s)&=\eta_{\sigma T_1}^yg_{\sigma}(s)\\
G_{\sigma}(x,y,s)&=\eta_{\sigma T_1}^y\eta_{\sigma T_2}^x\eta_{12}^{xy}g_{\sigma}(s),
\end{align}
where $g_\sigma (s)$ is defined as $G_\sigma (0,0,s)$. Equation \ref{eq_sig} then requires
\begin{align}
\eta_{\sigma}I &= (\eta_{\sigma T_1}\eta_{\sigma T_2})^{x+y}g_{\sigma}(\sigma(s))g_{\sigma}(s).
\end{align}
Because these relations must hold for all lattice positions $x$ and $y$, 
\begin{equation}
\eta_{\sigma T_1}\eta_{\sigma T_2} = 1
\end{equation}
and 
\begin{equation}
\label{eq_sigfin}
g_{\sigma}(\sigma(s))g_{\sigma}(s) = \eta_{\sigma}I
\end{equation}
(where $\sigma(u) = u, \sigma(v) = w$ and $\sigma(w) = v$). 

At this point, we look to add the rotational symmetry $C_3$. Before we solve the gauge matrix equations, we note that we have the freedom to multiply any of the gauge transformations by elements of the IGG, as a gauge transformation. Due to the fact that we are examining $\mbz_2$ spin liquids, we can make the transformation
\begin{align}
G_{T_2} &= \eta_{C_3 T_1}G_{T_2}, \\
G_{T_1} &= \eta_{C_3 T_1}\eta_{C_3 T_2}G_{T_1}, \\
G_{C_3} &= -\eta_{C_3} G_{C_3}
\end{align}
resulting in the fixing of the variables $\eta_{C_3 T_1}, \eta_{C_3 T_2}$ and $\eta_{C_3}$.

Next, we wish to write $G_{C_3}$ as a product of a term which depends only on sublattice and a term which depends only on lattice position, as we did for $G_\sigma$. We note that eq. \ref{eq_S1} and \ref{eq_S2} can be rewritten as 
\begin{align}
G_{C_3}^\dag(T_2^{-1}T_1(i))G_{C_3}(i) \eta_{12}^{x+1}&= I,\\
G_{C_3}^\dag(T_1(i))G_{C_3}(i)\eta_{12}^y &= I,
\end{align}
which, substituting $G_{C_3}(0,0,s) = -g_{C_3}(s)$, leads to
\begin{align}
G_{C_3}(n,-n,s) &= -\eta_{12}^{n(n-1)/2}g_{C_3}(s),\\
G_{C_3}(x,y,s) &= -\eta_{12}^{y(y+1)/2+y(x+y)}g_{C_3}(s).
\end{align}

Having the forms of $G_\sigma$ and $G_{C_3}$, we can consider the relations Eq. \ref{eq_sigS} and \ref{eq_S}. Eq. \ref{eq_sigS} simplifies to
\begin{align}
\eta_{\sigma C_3} I &= \eta_{\sigma T_1}^{y+1}g_{C_3}( \sigma(s)) g_\sigma(\sigma(s))g_{C_3}(s)g_\sigma(C_3^{-1}(s))
\end{align}
which tells us that $\eta_{\sigma T_1} = 1$, as both sides of this equation must be $y$ independent. Considering the different sublattice indices, we find
\begin{align}
\label{eq_c3sig1}
\eta_{\sigma C_3} I &= g_{C_3}(w) g_\sigma(u)g_{C_3}(u)g_\sigma(v) \\
&= g_{C_3}(v) g_\sigma(w)g_{C_3}(v)g_\sigma(w).
\label{eq_c3sig2}
\end{align}
Eq.  \ref{eq_S} simplifies to 
\begin{align}
\eta_{12} g_{C_3}(u)g_{C_3}(v)g_{C_3}(w) = I.
\label{eq_gc3}
\end{align}
It is then convenient to make the further gauge transformation $G_{C_3} \rightarrow \eta_{12} G_{C_3}$, which removes $\eta_{12}$ from Eq. \ref{eq_gc3}.

When considering the time reversal operation, it is convenient to consider bosonic and fermionic spin liquids separately. In the case of fermionic spin liquids, we consider the time reversal operator adjoined to a gauge transformation $i\tau_2$ rather than the bare operator (where $\vec{\tau}$ represents the vector of Pauli matrices). As $i\tau_2$ is a gauge transformation, it has no physical effect, but the ansatz now transforms as $T U_{ij} T^{-1} = -U_{ij}$. This combined operator satisfies $T^2 = e$ and $S^{-1}T^{-1}ST = e$. In addition, this operator acting on a pure gauge transformation satisfies $TGT^{-1} = G$, and can therefore be considered adjoined to a gauge transformation in the same fashion as we considered other space group symmetries,
\begin{gather}
[G_{T}(i)]^2 = \eta_{T} I, \label{eq_T}\\
G_{T_1}^\dag(i)G_{T}^\dag(i)G_{T_1}(i)G_{T}(T_1^{-1}(i)) = \eta_{T_1 T} I, \label{eq_1T}\\
G_{T_2}^\dag(i)G_{T}^\dag(i)G_{T_2}(i)G_{T}(T_2^{-1}(i)) = \eta_{T_2 T} I, \label{eq_2T}\\
G_{\sigma}^\dag(i)G_{T}^\dag(i)G_{\sigma}(i)G_{T}(\sigma^{-1}(i)) = \eta_{\sigma T} I, \label{eq_sigT}\\
G_{C_3}^\dag(i)G_{T}^\dag(i)G_{C_3}(i)G_{T}(C_3^{-1}(i)) = \eta_{C_3 T} I,\label{eq_ST}.
\end{gather}

Eq. \ref{eq_1T} and \ref{eq_2T} restrict $G_T(x,y,s) = \eta_{T_1 T}^x\eta_{T_2 T}^y g_T(s)$, where $g_T(s) = G_T(0,0,s)$. Eq. \ref{eq_sigT} then implies
\begin{align}
g_{\sigma}^\dag(s) g_T^\dag(s)g_\sigma(s)g_T(\sigma(s))\eta_{T_1 T}^{x+y}\eta_{T_2 T}^{x+y} = \eta_{\sigma T} I
\end{align}
leading to $\eta_{T_1 T}=\eta_{T_2 T}$. Considering next Eq. \ref{eq_ST}, we find 
\begin{align}
g_{C_3}^\dag(s) g_T^\dag(s)g_{C_3}(s)g_T(C_3^{-1}(s))\eta_{T_1 T}^{y+1} = \eta_{C_3 T} I
\end{align}
giving us also that $\eta_{T_1 T}= 1$. As such, $G_T(x,y,s) = g_T(s)$.

Because $T U_{ij} T^{-1} = -U_{ij}$, we require $U_{ij}=-G_T(i)U_{ij}G_T^\dagger(j)$. As the lattice is not bipartite and we require the ansatz to be non-zero on all bonds, this requires $G_T(x,y,s) = g_T(s) = i\vec{a}\cdot \vec{\tau}$, implying $\eta_T = 1$. We perform a sublattice dependent gauge transformation $W_s$ on each sublattice, such that $W_s g_T(s) W_s^\dag = i\tau_2$. We further note that $\eta_{C_3 T}$ must always be 1, as $g_{C_3}^\dag(s) \tau_2 g_{C_3}(s) \tau_2 = -1 ~ \forall s$ is incompatible with \ref{eq_gc3}. The remaining commutation equation simplifies to
\begin{equation}
\label{eq_sigTfin}
g_\sigma^\dag(s) \tau_2 g_\sigma(s) \tau_2  = \eta_{\sigma T} I.
\end{equation}

A sublattice dependent gauge transformation $W_s = e^{i \theta_s \tau_2}$ can be performed on each site, without affecting the previous results. This results in a change of the sublattice portions of the gauge transformations,
\begin{align}
g_{\sigma}(u) &\rightarrow W_u g_{\sigma}(u) W_u^\dag \label{eq_1}, \\
g_{\sigma}(v) &\rightarrow W_v g_{\sigma}(v) W_w^\dag \label{eq_2}, \\
g_{\sigma}(w) &\rightarrow W_w g_{\sigma}(w) W_v^\dag \label{eq_3}, \\
g_{C_3}(u) &\rightarrow W_u g_{C_3}(u) W_v^\dag \label{eq_4},\\
g_{C_3}(v) &\rightarrow W_v g_{C_3}(v) W_w^\dag \label{eq_5},\\
g_{C_3}(w) &\rightarrow W_w g_{C_3}(w) W_u^\dag \label{eq_6}. 
\end{align}

We must now solve the equations \ref{eq_sigfin}, \ref{eq_c3sig1}, \ref{eq_c3sig2}, \ref{eq_gc3} and \ref{eq_sigTfin}. We note that $\eta_{C_3 T} = 1 \Rightarrow g_{C_3}(s) = e^{i\tau_2 \theta_s}$, with $\theta_s \in [0,2\pi)$. Using the gauge transformations $W_u = e^{-i\theta_u \tau_2}$ and $W_w = e^{i\theta_v \tau_2}$ results in a fixing $g_{C_3}(s) = I$.

Manipulating Eq. \ref{eq_sigfin}, \ref{eq_c3sig1} and \ref{eq_c3sig2} with $g_{C_3}(s) = I$ we see that $g_\sigma(u) = g_\sigma(w)$. This further implies that $\eta_\sigma = \eta_{\sigma C_3}$, and therefore $g_\sigma(v) = g_\sigma(u)$. 

Different solutions arise for $\eta_{\sigma T} = \pm 1$. When $\eta_{\sigma T} = 1$, eq. \ref{eq_sigTfin} gives $g_\sigma(s) = e^{i\tau_2 \phi}$ for some $\phi \in [0,2\pi)$. This is further restricted by \ref{eq_sigfin} to $g_\sigma(s) = \pm I$ for $\eta_\sigma = 1$ or $g_\sigma(s) = \pm i\tau_2$ for $\eta_\sigma = -1$. Finally, we can choose the positive solution for each of these, by performing the gauge transformation $G_\sigma \rightarrow \pm G_\sigma$. When $\eta_{\sigma T} = -1$, eq. \ref{eq_sigTfin} requires $g_\sigma(s) = i\tau_3e^{i\tau_2 \phi}$ for some $\phi \in [0,2\pi)$. Using the gauge transformation $W_u = W_v = W_w = e^{i\tau_2 \phi/2}$, we can fix $g_\sigma(s) = i\tau_3$. This gives a total of three solutions to the sublattice dependent equations, each of which can have $\eta_{12} = \pm 1$. Therefore, we have six PSGs, the details of which are summarized in table \ref{tab_psgsol}.

Next we describe the bosonic PSG. All of the solution prior to the consideration of time reversal is equivalent for bosons. In the bosonic case, we can solve the equations \ref{eq_sigfin}, \ref{eq_c3sig1}, \ref{eq_c3sig2} and \ref{eq_gc3} directly, and show that these solutions are compatible with the time reversal symmetry.

First, we note that the sublattice dependent gauge transformation $W_s = e^{i \theta_s}$ is allowed for bosons, which have the same effect (Eq. \ref{eq_1} - \ref{eq_6}) as for the fermionic operators. Denoting $g_S(s) = e^{i\phi_S(s)}$ and $\eta_\alpha = e^{i n_\alpha}$, we see from \ref{eq_sigfin} that $\phi_\sigma(u) = n_\sigma \pi /2$. Further, performing a gauge transformation $\theta_v = (\phi_\sigma(w) - \phi_\sigma(v))/2$, we can fix $\phi_\sigma(v) = \phi_\sigma(w) = \phi_\sigma(u)= n_\sigma \pi /2$.

Next, we see from eq. \ref{eq_c3sig1} and \ref{eq_c3sig2} that $\phi_3(w) + \phi_3(u) = (n_{\sigma C_3} - n_{\sigma})\pi$ and $\phi_3(v) = (n_{\sigma C_3} - n_{\sigma})\pi/2$. Combined with eq. \ref{eq_gc3}, we see that $n_\sigma = n_{\sigma C_3}$, and therefore we can choose $\phi_3(v) = 0$ and $\phi_3(u)=\phi_3(w)$. Performing a gauge transformation $\theta_u = -\phi_3(u)$, we arrive at $\phi_3(u)=\phi_3(v)=\phi_3(w)=0$.

Finally, we consider the effect of time reversal for bosons. Due to the condition $\phi_3(s) = 0$, all of the bonds on each type of triangle must have the same complex phase. Using an overall gauge transformation on each site on the lattice, we can enforce the condition that these are all real on one type of triangle. In this gauge, it is clear that time reversal acts non-projectively. Thus, an ansatz satisfies the time reversal symmetry iff the other triangle is also real in this gauge.

\section{\label{app:vmc}Details of the numerical calculation}

The SR method is described in detail by Yunoki {\it et. al.}\cite{yunoki2006two}. We outline the method here, for completeness. The overlap of the mean field wavefunction and a given physical spin state $|x\rangle$ takes the form $\Psi_{\alpha_k}(x) = \langle x | \Psi_{\alpha_k} \rangle$. Further, as $|x\rangle$ is a physical state, $P|x\rangle = |x\rangle$, and therefore $\Psi_{\alpha_k}(x)$ is also the overlap of the projected wavefunction with the spin state. The change of the wavefunction due to a small change in the parameters $\alpha_k \rightarrow \alpha'_k = \alpha_k + \delta\alpha_k$ is
\begin{equation}
\Psi_{\alpha'_k}(x) = \Psi_{\alpha_k}(x)\left[1+\sum_k O_k(x)\delta\alpha_k + O(\delta\alpha_k^2)\right],
\end{equation}
where $O_k(x)$ is defined as the logarithmic derivitive of the wavefunction with respect to the variational parameter $\alpha_k$. These $O_k(x)$ can be determined for a particular value of the parameters $\alpha_k$ and a particular spin state $|x\rangle$ through a matrix multiplication. It is also convenient to define the operator $\hat{O}_k$, with the property $\langle x|\hat{O}_k|x'\rangle = O_k(x)\delta_{xx'}$.

The expectation value of an operator $Q$ in a state $|\Psi \rangle$ can be calculated using 
\begin{align}
\langle Q \rangle_{\Psi} &= \sum_x P(x) \frac{\langle x | Q | \Psi \rangle}{\langle x | \Psi \rangle}, \\
P(x) &\equiv \frac{\langle \Psi | x \rangle \langle x| \Psi \rangle }{\sum_{x'} \langle \Psi | x' \rangle \langle x'| \Psi \rangle}.
\end{align}
Considering $P(x)$ as a probability distribution, we can sample the physical spin states to determine this using a Monte Carlo sampling. In particular, we can derive both the energy and the derivatives of the energy with respect to the variational parameters (generalized forces) in this fashion, as
\begin{align}
E(&\Psi_{\alpha_k}) = \langle \hat{H} \rangle_{\Psi_{\alpha_k}}, \\
f_k &\equiv -\frac{\partial E(\Psi_{\alpha_k})}{\partial \alpha_k} = -2\langle \hat{H}\hat{O}_k  \rangle_{\Psi_{\alpha_k}} + 2 \langle \hat{H} \rangle_{\Psi_{\alpha_k}} \langle \hat{O}_k \rangle_{\Psi_{\alpha_k}},
\end{align}
where we have assumed $O_k(x)$ is real, as is the case for our calculation.

Having computed the generalized forces, we can attempt to find the variational parameters which lead to a minimum of the energy landscape using the steepest descent method, ie. iteratively changing our parameters according to $\alpha_k \rightarrow \alpha'_k = \alpha_k + f_k\cdot \delta t$, where $\delta t$ can be determined at each step, or fixed to a sufficiently small value. However, this method is still effective if the forces are instead multiplied by a positive definite matrix $s$, such that
\begin{equation}
\alpha_k \rightarrow \alpha'_k = \alpha_k + \delta t \sum_l s_{k,l}^{-1} f_l.
\end{equation}
It turns out that a matrix
\begin{equation}
s_{i,j} = \langle \hat{O}_i \hat{O}_j \rangle_{\Psi_{\alpha_k}} - \langle \hat{O}_i \rangle_{\Psi_{\alpha_k}} \langle \hat{O}_j \rangle_{\Psi_{\alpha_k}},
\end{equation}
is more appropriate for this calculation. This is due to the fact that a small change in the variational parameters can correspond to a large change in the distance between the two normalized wavefunctions, i.e.
\begin{equation}
\Delta_\alpha = 2-2\frac{ \langle \Psi_{\alpha_k} | \Psi_{\alpha'_k} \rangle }{\sqrt{\langle \Psi_{\alpha_k} | \Psi_{\alpha_k} \rangle \langle \Psi_{\alpha'_k} | \Psi_{\alpha'_k} \rangle } }.
\end{equation}
By using this choice of $s$, at each step in the SR minimization we move small amounts in the wavefunction distance, rather than in the variational parameter distance.

\end{document}